\documentclass{optica-article}

\journal{opticajournal} 

\articletype{Research Article}

\usepackage{lineno}

\begin{document}

\title{Operating Single-Photon Circulator by Spinning Optical Resonators}
\author{Jing Li,\authormark{1,2} Tian-Xiang Lu,\authormark{3} Meiyu Peng, %
\authormark{1,2} Le-Man Kuang,\authormark{1,2} Hui Jing\authormark{1,2,*} and Lan Zhou\authormark{1,2,$\dagger$}}
\email{ \authormark{*} jinghui73@foxmail.com ,\authormark{$\dagger$}zhoulan@hunnu.edu.cn}

\address{\authormark{1} Key Laboratory of Low-Dimension Quantum Structures and Quantum Control of Ministry of Education, Key Laboratory for Matter Microstructure and Function of Hunan Province, Synergetic Innovation Center for Quantum Effects and Applications, Xiangjiang-Laboratory and Department of Physics, Hunan Normal University, Changsha 410081, China \\
\authormark{2}Institute of Interdisciplinary Studies, Hunan Normal University, Changsha, 410081, China \\
\authormark{3}College of Physics and Electronic Information, Gannan Normal University, Ganzhou 341000, Jiangxi, China\\
	}



\begin{abstract*}
A circulator is one of the crucial devices in quantum networks and simulations. We propose a four-port circulator that regulate the flow of single photons at muti-frequency points by studying the coherent transmission of a single photon in a coupled system of two resonators and two waveguides. When both resonators are static or rotate at the same angular velocity, single-photon transport demonstrates reciprocity; however, when the angular velocities differ, four distinct frequency points emerge where photon circulation can occur. In particular, when the angular velocities of the two resonators are equal and opposite, there are two different frequency points where photon circulation can be achieved, and there is a frequency point where a single photon input from any waveguide can be completely routed to the other waveguide. Interestingly, by rotating the two resonators, the single-photon circulation suppressed by the internal defect-induced backscattering can be restored.
		
\end{abstract*}


\section{Introduction}

Nonreciprocal optical devices, such as isolators \cite%
{xia19,Jalas13,Stadler14}, circulators \cite{Ruesink18,Navarathna23,Xu23},
and directional amplifiers \cite{jiang18}, featuring different optical
responses after exchanging positions between the input and output terminals,
are not only important components in optical systems but also have important
applications in constructing quantum networks and implementing quantum
communication. One of the most basic requirements for achieving
nonreciprocal transmission in optical systems is to break the symmetry of
time inversion \cite{Aplet64}. In recent years, in order to meet the
requirements of chip integration, a large number of non-magnetic and
nonreciprocal systems have begun to develop, which are generally based on
dynamic spatiotemporal modulation structures \cite{Qiu22}, quantum Hall
effects \cite{Viola14}, Kerr-nonlinear microresonators \cite%
{Fan12,Hua16,Cao17,Zhang19,Yang23}, optomechanical resonator \cite%
{Manipatruni09,Hafezi12,Habraken12,Schmidt15,
Metelmann15,Xu15,Xu16,Shen16,Ruesink16,Verhagen17,
Fang17,Peterson17,Bernier17,Barzanjeh17,Tian17,Jiang17,Xu19,
Mercier19,Lai20,Xu20,Chen21,Liu23,Lu2312,Zhao20,Zheng24,Zhang24},
non-Hermitian systems \cite%
{Xin15,Wu14,Rter10,Bender13,Peng14,Chang14,Chang20,Wang20,Wang2013}, moving
atomic gases \cite{Horsley13,Ramezani18,Zhang18,Xia18,Lin19,Liang20,Li20,
Hu21}, and spinning resonators \cite{Maayani18,Jing2018,Graf22}.

Recent experimental studies \cite{Maayani18,Li2019,Mao2022,Khia2018} have
shown that spinning resonators can break the time reversal symmetry via
Fizeau resistance to achieve nonreciprocal transmission of light.
Afterwards, more and more researchers became interested in rotating
resonators and proposed many nonreciprocal quantum devices, such as
nonreciprocal photon blocking \cite%
{Huang18,Huang19,Wang2019,Shen2020,Xue2020,Jing2021}, nonreciprocal phonon
lasers \cite{JiangY2018,Xu21}, nonreciprocal entanglement \cite{Jiao20},
nonreciprocal optical solitons \cite{Li21}, nonreciprocal optical bandwidth
\cite{Hu23,Tang22}, and photon circulators \cite{wei20}. Circulator is
typically a nonreciprocal device composed of three or four ports. They
separate the paths of the sender and receiver by introducing non reciprocity
between ports. The circulator has been experimentally proven by microwave
\cite{Kerckhof15,Sliwa15,Chapman17,Muller18} and optical signals \cite%
{Shen18,Scheucher16}, and can be applied in regions ranging from classical
to quantum \cite{Scheucher16,Chen17,Liu17,xu2020}. Recently, a two-frequency
point photon circulator based on a rotating resonator with two tapered
fibers has been proposed~\cite{wei20}. However, the novel possibility of a
four-frequency points single-photon circulator by utilizing two spinning
resonators, as far as we know, has not been explored.

In this paper, we found that four frequency points can achieve photon
circulator by utilizing two spinning resonators with different angular
velocities. To pursue this, we has been conducted coupling two rotating
resonators with two waveguides to form a four-port device. There are various
combinations of angular velocities for the two resonators, each resulting in
different outcomes. When the resonators are static or rotate at the same
angular velocity, photon transmission is reciprocal. However, with differing
angular velocities, it's been observed that four frequency points of photon
circulation can be achieved. Particularly, when the two resonators have
equal and opposite angular velocities, two frequency points of photon
circulation are attainable. Additionally, there exists a frequency point
where photons input from any waveguide can be completely routed to another.
We also found that, compared with other nonreciprocal device
circulators, single-photon circulators formed by rotating resonators
exhibits robustness against the backscattering. Therefore, the routing
direction of the photon circulator is dependent on the frequency of incoming
photons and the rotational angular velocities of the two resonators.
Compared to nonreciprocal devices with two ports, the four-port quantum
optical circulator with multi-frequency points enables us to establish
two-dimensional and three-dimensional networks to achieve photon quantum simulations\cite{Georgescu14}. Our results may provide inspiration for multi-frequency points circulator in quantum information.

The paper is organized as follows: In Sec.~2, we outline the theoretical
model of the system by presenting Hamiltonian. We give the calculation
method and solution to the model. In Sec.~3, we study the single-photon
transport properties when two resonators rotate or static. In Sec.~4, we
study the effect of adding backscattering in resonators on photon
transmission. Finally, we summarize the single-photon transport and give the
conclusions in Sec.~5.


\section{\label{Sec:2} Theoretical model}

Recent experiments have demonstrated the utilization of spinning devices for
achieving optical isolators\cite{Maayani18}, one-way heat flow~\cite%
{Li2019,Xu2020}, gyroscopes\cite{Mao2022,Khia2018}, acoustic amplifiers\cite{Cromb2020},
and rotational doppler effect\cite{Chuan24}. Among these, utilizing a spinning resonator can achieve
nonreciprocal transmission with an isolation degree of up to 99.6$\%$\cite%
{Maayani18}. In this experiment, a spherical resonator was successfully
created by flame polishing the end of a molten silicon glass cylinder and
mounting the resonator on a turbine. By properly placing a fiber near the
spinning resonator, a stable fiber-resonator coupling was established
through a "self-adjusting" aerodynamic process.

In this paper, we propose how to realize a single-photon circulator with
multiple frequency points in a system composed of two spinning
whispering-gallery-mode (WGM) resonators and two waveguides. As shown in
Fig.~1(a), we consider two spinning WGM microresonators $f$ and $d$, coupled
to two waveguides a and b, with four input/output ports, respectively. The
WGM holds two cavity modes corresponding to clockwise (CW) and
counterclockwise (CCW) modes with frequencies $\omega_{cw}=\omega_{c}-%
\Delta_{F}$ and $\omega_{ccw}=\omega_{c}+\Delta_{F}$, respectively. When the
resonator spins at an angular velocity $\Omega$, the rotation-induced
Sagnac-Fizeau shift $\Delta_{F}$ is given by \cite{Maayani18,Malykin2000}
\begin{eqnarray}
\Delta _{F}=\frac{nR\Omega \omega _{c}}{c}\left( 1-\frac{1}{n^{2}}-\frac{
\lambda }{n}\frac{dn}{d\lambda }\right)= \Omega G,  \label{1-1}
\end{eqnarray}
where $\omega_{c}$ is the intrinsic frequency of the nonspinning resonator, $%
c$ and $\lambda$ are the propagation speed and wavelength of light, $R$ and $%
n$ are the radius and reflectivity of the resonator, respectively. The
dispersion term $dn/{d\lambda }$, characterizing the relativistic origin of
the Sagnac shift, is relatively small in typical materials ($\sim1\%$) \cite%
{Malykin2000}. $\Omega>0$ ($\Omega<0$) indicates CW (CCW) rotation of the
resonator.

$\Omega_{1}$ ($\Omega_{2}$) is the rotational angular velocity of the $f$ ($d
$) resonator, and $\Delta_{F1}$ ($\Delta_{F2}$) is the Sagnac shift due to
the rotation of the $f$ ($d$) resonator.  Under the rotating wave
approximation, the total Hamiltonian can be written as ($\hbar=1$)
\begin{eqnarray}
\hat{H}=\hat{H}_{0}+\hat{H}_{\mathrm{int}} .  \label{1-2}
\end{eqnarray}
Here, the first term $\hat{H}_{0}$ describes the free part,
\begin{eqnarray}
\hat{H}_{0} &=&\sum_{\alpha =cw,ccw}\left( \omega _{\alpha }\hat{f}_{\alpha
}^{\dagger }\hat{f}_{\alpha }+\omega _{d,\alpha }\hat{d}_{\alpha }^{\dagger
} \hat{d}_{\alpha }\right) -iv_{g}\int dx\hat{a}_{R_{x}}^{\dagger }\frac{%
\partial }{\partial x}\hat{a} _{R_{x}}+iv_{g}\int dx\hat{a}_{L_{x}}^{\dagger
}\frac{\partial }{\partial x} \hat{a}_{L_{x}}  \nonumber \\
&&-iv_{g}\int dx\hat{b}_{R_{x}}^{\dagger }\frac{\partial }{\partial x}\hat{b}
_{R_{x}}+iv_{g}\int dx\hat{b}_{L_{x}}^{\dagger }\frac{\partial }{\partial x}
\hat{b}_{L_{x}},  \label{1-3}
\end{eqnarray}
where $\omega_{\alpha}=\omega_c\pm\Delta_{F1}$ ($\omega_{d,\alpha}=\omega_c%
\pm\Delta_{F2}$) is the effective frequency for $\alpha$ $(\alpha=cw,ccw)$
travelling modes in the $f$ ($d$) resonator. $\hat{f}_{\alpha}^{\dagger}$ ($%
\hat{d}_{\alpha}^{\dagger}$) is the creation operators in the $f$ ($d$%
)resonator. $\hat{a}^{\dagger}_{R_{x}}$  ($\hat{a}^{\dagger}_{L_{x}}$) is the creation operators for the right-moving
(left-moving) photon along the waveguide $a$ at position $x$. $\hat{b}^{\dagger}_{R_{x}}$
  ($\hat{b}^{\dagger}_{L_{x}}$) is the creation operators for the right-moving
(left-moving) photon along the waveguide $(b)$ at position $x$. The
group velocities have been assumed to be $v_{g}$.  The second terms $\hat{H}%
_{int}$ in Eq.~(\ref{1-1}) describes the interacting part.
\begin{eqnarray}
\hat{H}_{\mathrm{int}} &=&g_{a}(\hat{a}_{R_{0}}^{\dagger }\hat{f}_{cw}+\hat{a}
_{L_{0}}^{\dagger }\hat{f}_{ccw})+g_{b}(\hat{b}_{L_{0}}^{\dagger }\hat{d}
_{cw}+\hat{b}_{R_{0}}^{\dagger }\hat{d}_{ccw})  \nonumber \\
&&+J(\hat{f}_{cw}^{\dagger }\hat{d}_{ccw}+\hat{f}_{ccw}^{\dagger }\hat{d}
_{cw})+\mathrm{H.c.}  \label{1-4}
\end{eqnarray}
where $g_{a}$ ($g_{b}$) is the $f$ ($d$) resonator and $a$ ($b$) waveguide
coupling strength. $J$ is the coupling strength of resonator $f$ and $d$.

\begin{figure}[tbp]
\centering
\includegraphics[width=0.95\textwidth]{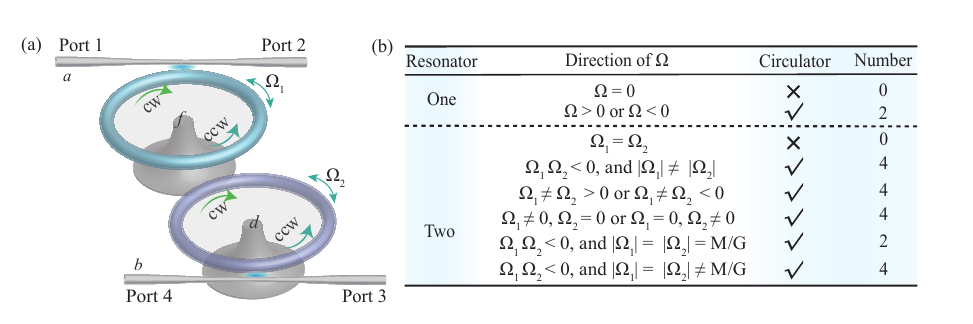}
\caption{(a) Schematic of the system under study. We consider two rotating
WGM microresonator coupled to two waveguides with four input/output ports,
respectively. (b) A list of results for single and two resonators. The symbol $\times$
indicates that there is no frequency point to achieve a photon circulator.
The symbol $\surd$ indicates that there is frequency point to achieve a
photon circulator. $M=%
\protect\sqrt{g_{a}^{2}g_{b}^{2}+4v_g^{2}J^{2}}/(2v_{g})$, $G=R \protect%
\omega _{c}\left( n^{2}-1\right)/(cn)$ in the table.}
\label{fig-1.pdf}
\end{figure}
\vspace{0pt}

In the case of a single excitation, the state vector of the system can be
written as
\begin{eqnarray}
\left\vert \psi \right\rangle &=&\sum_{\alpha =cw,ccw}\left( C_{\alpha }\hat{%
f} _{\alpha }^{\dagger }\left\vert \varnothing \right\rangle +D_{\alpha }%
\hat{d} _{\alpha }^{\dagger }\left\vert \varnothing \right\rangle \right)
\nonumber \\
&&+\sum_{\beta =L,R}\left( \int dxA_{\beta }\left( x\right) \hat{a}_{\beta
_{x}}^{\dagger }+\int dxB_{\beta }\left( x\right) \hat{b}_{\beta
_{x}}^{\dagger }\right) \left\vert \varnothing \right\rangle,  \label{1-5}
\end{eqnarray}
where $\left\vert \varnothing \right\rangle$ is the vacuum state which
indicates that there is zero photon in both the waveguide and the resonator.
$C_{\alpha}$ ($D_{\alpha}$) is the probability amplitude of a photon
appearing in $f$ ($d$) resonator. $A_{\beta}(x)$ ($B_{\beta}(x)$, $\beta=L, R
$) is the probability amplitude of the right- or left-moving photon in the
waveguide a (b). According to the Schr$\ddot{o}$dinger equation $\hat{H}%
\left\vert \psi \right\rangle=E\left\vert \psi \right\rangle$, and removing $%
C_{\alpha}$ and $D_{\alpha}$, the coupled equations can be written

\begin{eqnarray}
\begin{aligned} \label{1-7} EA_{L}\left( x\right) -iv_{g}\frac{\partial
A_{L}\left( x\right) }{\partial x} &=\delta \left( x\right)
\frac{g_{a}^{2}\left( E-\omega _{d,cw}\right) A_{L}\left( 0\right)
+Jg_{a}g_{b}B_{L}\left( 0\right) }{\left( E-\omega _{ccw}\right) \left(
E-\omega _{d,cw}\right) -J^{2}}, \\ EB_{L}\left( x\right)
-iv_{g}\frac{\partial B_{L}\left( x\right) }{\partial x} &=\delta \left(
x\right) \frac{g_{b}^{2}\left( E-\omega _{ccw}\right) B_{L}\left( 0\right)
+Jg_{a}g_{b}A_{L}\left( 0\right) }{\left( E-\omega _{ccw}\right) \left(
E-\omega _{d,cw}\right) -J^{2}}, \\ EA_{R}\left( x\right)
+iv_{g}\frac{\partial A_{R}\left( x\right) }{\partial x} &=\delta \left(
x\right) \frac{g_{a}^{2}\left( E-\omega _{d,cw}\right) A_{R}\left( 0\right)
+Jg_{a}g_{b}B_{R}\left( 0\right) }{\left( E-\omega _{cw}\right) \left(
E-\omega _{d,ccw}\right) -J^{2}}, \\ EB_{R}\left( x\right)
+iv_{g}\frac{\partial B_{R}\left( x\right) }{\partial x } &=\delta \left(
x\right) \frac{g_{b}^{2}\left( E-\omega _{cw}\right) B_{R}\left( 0\right)
+Jg_{a}g_{b}A_{R}\left( 0\right) }{\left( E-\omega _{cw}\right) \left(
E-\omega _{d,ccw}\right) -J^{2}}. \end{aligned}
\end{eqnarray}
The photon enters through port 1 of the $a$ waveguide, and the wave function
is
\begin{eqnarray}
\begin{aligned} \label{1-8} A_{R}\left( x\right)
&=e^{i\frac{E}{v_{g}}x}\left( \theta \left( -x\right) +t_{1\rightarrow
2}\theta \left( x\right) \right), \\ A_{L}\left( x\right)
&=e^{-i\frac{E}{v_{g}}x}t_{1\rightarrow 1}\theta \left( -x\right), \\
B_{R}\left( x\right) &=e^{i\frac{E}{v_{g}}x}t_{1\rightarrow 3}\theta \left(
x\right), \\ B_{L}\left( x\right) &=e^{-i\frac{E}{v_{g}}x}t_{1\rightarrow
4}\theta \left( -x\right), \end{aligned}
\end{eqnarray}
where $t_{i\rightarrow j}$ $(i,j={1,2,3,4})$ represents the probability
amplitude of photon transmission from port $i$ to port $j$. $\theta\left( \pm
x\right)$ is a step function. The transmission probability is defined by
\begin{eqnarray}
T_{i\rightarrow j}=|t_{i\rightarrow j}|^{2}.
\end{eqnarray}
Combine Eq.~(\ref{1-7}) and Eq.~(\ref{1-8}), we obtain the transmission
probability amplitude
\begin{eqnarray}
t_{1\rightarrow 1} &=&0,\qquad t_{1\rightarrow 4}=0, \\
t_{1\rightarrow 2} &=&\frac{\left( \delta +\Delta _{F1}-i\frac{g_{a}^{2}}{
2v_{g}}\right) \left( \delta -\Delta _{F2}+i\frac{g_{b}^{2}}{2v_{g}}\right)
-J^{2}}{\left( \delta +\Delta _{F1}+i\frac{g_{a}^{2}}{2v_{g}}\right) \left(
\delta -\Delta _{F2}+i\frac{g_{b}^{2}}{2v_{g}}\right) -J^{2}} ,  \label{1-9}
\\
t_{1\rightarrow 3} &=&\frac{-i\frac{1}{v_{g}}Jg_{a}g_{b}}{\left( \delta
+\Delta _{F1}+i\frac{g_{a}^{2}}{2v_{g}}\right) \left( \delta -\Delta _{F2}+i
\frac{g_{b}^{2}}{2v_{g}}\right) -J^{2}} ,  \label{1-10}
\end{eqnarray}
where $\delta=E-\omega_{c}$ is detuning. Since the direction of photon
transmission is opposite to the direction of the CCW mode in $f$ resonator, $%
T_{1\rightarrow 1}=0$. Similarly, since the CW mode of the $f$ resonator is
only coupled to the CCW mode of the $d$ resonator, $T_{1\rightarrow 4}=0$.
So, it is easily to find that $T_{1\rightarrow 2}+T_{1\rightarrow 3}=1$
which guarantees the probability conservation for the incident photon.

If the photon enters through port 2 of the $a$ waveguide, and the wave
function is
\begin{eqnarray}
\begin{aligned} \label{1-11} A_{L}\left( x\right)
&=e^{-i\frac{E}{v_{g}}x}\left( \theta \left( x\right) +t_{2\rightarrow
1}\theta \left( -x\right) \right), \\ A_{R}\left( x\right)
&=e^{i\frac{E}{v_{g}}x}t_{2\rightarrow 2}\theta \left( x\right), \\
B_{R}\left( x\right) &=e^{i\frac{E}{v_{g}}x}t_{2\rightarrow 3}\theta \left(
x\right), \\ B_{L}\left( x\right) &=e^{-i\frac{E}{v_{g}}x}t_{2\rightarrow
4}\theta \left( -x\right). \end{aligned}
\end{eqnarray}

The probability amplitude can be obtained by combining Eqs.~(\ref{1-7}) and (%
\ref{1-11})
\begin{eqnarray}
t_{2\rightarrow 2} &=&0,\qquad t_{2\rightarrow 3}=0, \\
t_{2\rightarrow 1} &=&\frac{\left( \delta -\Delta _{F1}-i\frac{g_{a}^{2}}{
2v_{g}}\right) \left( \delta +\Delta _{F2}+i\frac{g_{b}^{2}}{2v_{g}}\right)
-J^{2}}{\left( \delta -\Delta _{F1}+i\frac{g_{a}^{2}}{2v_{g}}\right) \left(
\delta +\Delta _{F2}+i\frac{g_{b}^{2}}{2v_{g}}\right) -J^{2}} ,  \label{1-12}
\\
t_{2\rightarrow 4} &=&\frac{-i\frac{1}{v_{g}}\lambda g_{a}g_{b}}{\left(
\delta -\Delta _{F1}+i\frac{g_{a}^{2}}{2v_{g}}\right) \left( \delta +\Delta
_{F2}+i\frac{g_{b}^{2}}{2v_{g}}\right) -J^{2}}.  \label{1-13}
\end{eqnarray}
Using the same method, we can also obtain transmission probabilities $%
T_{3\rightarrow 4}$, $T_{3\rightarrow 1}$, $T_{4\rightarrow 3}$ and $%
T_{4\rightarrow 2}$. They satisfy the relationships $T_{1\rightarrow
2}=T_{4\rightarrow 3}$, $T_{1\rightarrow 3}=T_{4\rightarrow 2}$, $%
T_{2\rightarrow 1}=T_{3\rightarrow 4}$ and $T_{2\rightarrow
4}=T_{3\rightarrow 1}$. From Eqs.~(\ref{1-9}-\ref{1-10}) and (\ref{1-12}-\ref%
{1-13}), it can be observed that if $\Delta_{F1}=\Delta_{F2}$ and $g_{a}=g_{b}$, we can
observe that $T_{1\rightarrow 2}=T_{2\rightarrow 1}$, $T_{1\rightarrow
3}=T_{2\rightarrow 4}$. That is to say, when two resonators are static ($%
\Omega_{1}=\Omega_{2}=0$) or rotate with the same angular velocity ($%
\Omega_{1}=\Omega_{2}\neq0$), the photon transmission in the same
waveguide is reciprocal.
\begin{figure}[tbp]
\centering\includegraphics[width=0.97\textwidth]{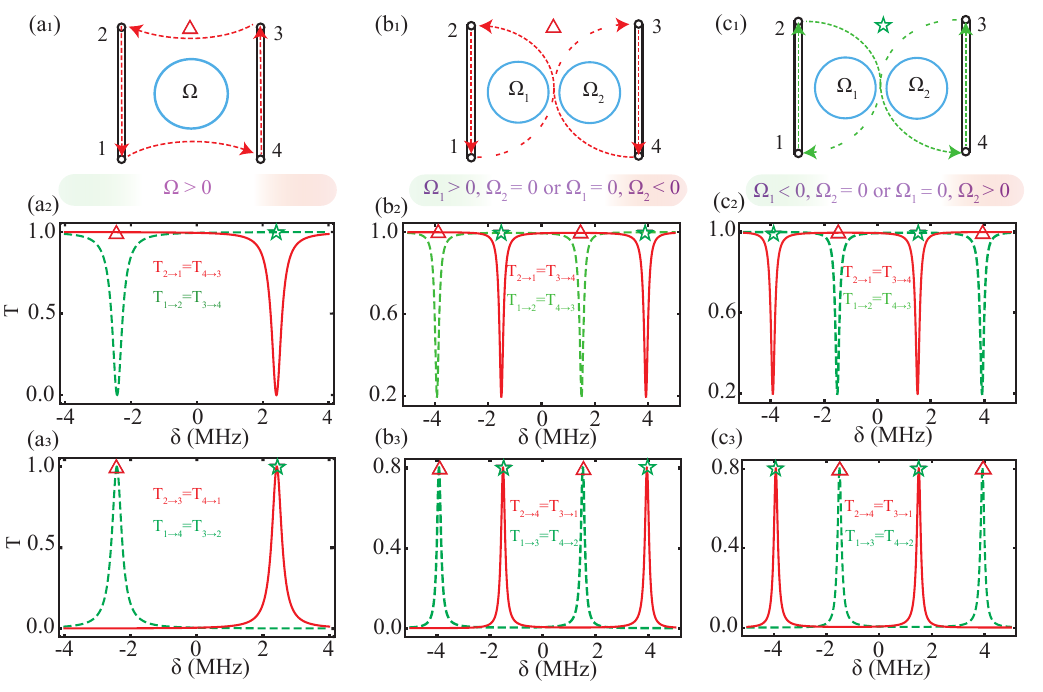}
\caption{The routing behaviors of the circulator for a single resonator ($\mathrm{a_1}$)
and two resonators ($\mathrm{b_1}$), ($\mathrm{c_1}$). In ($\mathrm{a_1}$), the direction of the circulator is $2\rightarrow 1\rightarrow 4 \rightarrow3 \rightarrow2$. In ($\mathrm{b_1}$), the direction of the circulator is $2\rightarrow 1 \rightarrow3 \rightarrow4 \rightarrow2$. In ($\mathrm{c_1}$), the direction of the circulator is $1\rightarrow 2\rightarrow 4 \rightarrow3 \rightarrow1$ . Transmission probability distribution as a
function of detuning $\protect\delta$ for a single resonator ($\mathrm{a_2}$-$\mathrm{a_3}$) and two
resonators ($\mathrm{b_2}$-$\mathrm{b_3}$), ($\mathrm{c_2}$-$\mathrm{c_3}$). ($\mathrm{a_2}$-$\mathrm{a_3}$) $\Omega=29$ kHz. ($\mathrm{b_2}$-$\mathrm{b_3}$) $\Omega_1=29$ kHz, $\Omega_2=0$.
($\mathrm{c_2}$-$\mathrm{c_3}$)  $\Omega_1=-29$ kHz, $\Omega_2=0$. Other parameters can be found in the main
text. }
\label{fig2}
\end{figure}


\section{\label{Sec:3} A Single-Photon Circulator Made of Two Spinning Optical Resonators}

\begin{figure}[tbp]
	\centering\includegraphics[width=0.9\textwidth]{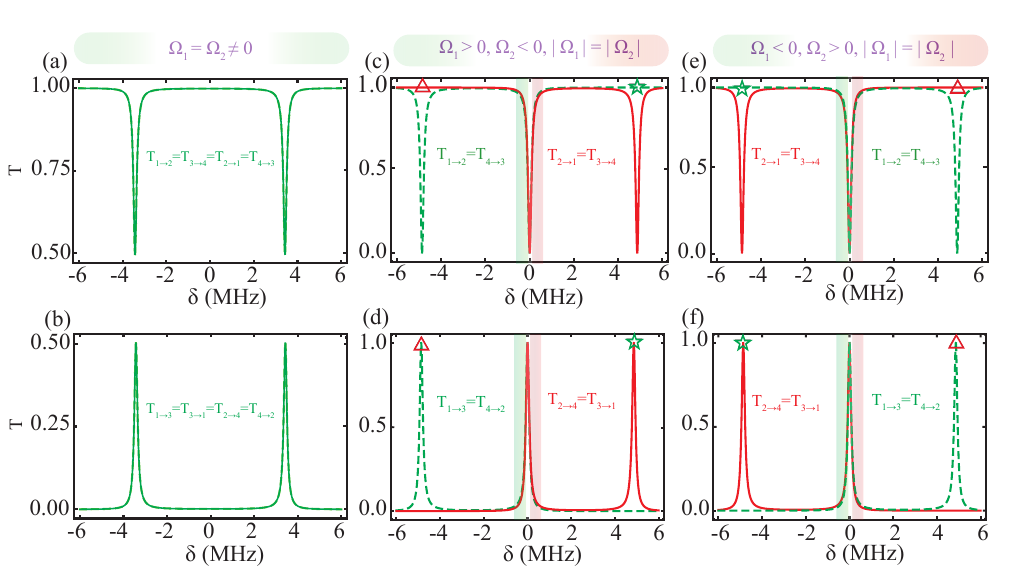}
	\caption{ Transmission probability distribution as a function of detuning $%
		\protect\delta$ for two resonators. (a-b) $\Omega_1=\Omega_2=29$ kHz. (c-d)
		$\Omega_1=29.2$ kHz, $\Omega_2=-29.2$ kHz. (e-f) $%
		\Omega_1=-29.2$ kHz, $\Omega_2=29.2$ kHz. Other parameters can
		be found in the main text.}
	\label{fig3}
\end{figure}
In this section, the main focus of these results is to analyze how different
spinning direction in the resonator can impact the frequency of the realized
photon circulator. For this purpose, we will focus on the following cases:
1) One resonator spins while the other remains static. 2) Both resonators
spin in the same direction. 3) Both resonators spin in opposite directions.
In our calculations, we have selected the  experimentally feasible
parameters \cite{Kerry03, Spillane05, Takao06}: $R=30$ $\mu$m, $n=1.4$%
, $Q=10^{9}$, $\lambda=1.55\times10^{-6}$ m, $\omega_c=2\pi c/\lambda$, $%
g_{a}=g_{b}=6\omega_c/Q$, $c=v_{g}=3\times 10^{8}$ m/s, and $J=2.4$ MHz.

For comparisons, we first consider the single resonator case illustrated in
Fig.\,\ref{fig2}($\mathrm{a_1}$). In Figs.\,\ref{fig2}($\mathrm{a_2}$) and \ref{fig2}($\mathrm{a_3}$), transmission
probability distribution as a function of detuning $\delta$ for single
resonator. The photon transmitted with circulation in the counterclockwise
direction $1\rightarrow 4\rightarrow 3\rightarrow 2\rightarrow 1$, at $%
\delta=\Delta_{F1}$ [see Fig.~2($\mathrm{a_1}$)]. In addition, the photon also can
transmit with circulation in the clockwise direction $1\rightarrow
2\rightarrow 3\rightarrow 4\rightarrow 1$, at $\delta=-\Delta_{F1}$. The
above circulation can be seen from the transmission spectra as shown in
Fig.~2($\mathrm{a_2}$-$\mathrm{a_3}$). The results are consistent with those in Ref.~\cite{wei20}.

For two resonators case with $\Omega_{1}>0$ and $\Omega_{2}=0$ (or $%
\Omega_{1}=0$ and $\Omega_{2}<0$), when $\delta =\pm \sqrt{%
g_{a}^{2}g_{b}^{2}+4v_{g}^{2}J^{2}+v_{g}^{2} \Delta _{F1} ^{2}}%
/(2v_{g})-\Delta _{F1}/2$, the photon transmitted with circulation in the
reverse $8$-word direction $2\rightarrow 1\rightarrow 3\rightarrow
4\rightarrow 2$, see Fig.~2($\mathrm{b_1}$) (The position marked by a red triangle). The
circulation can be seen from the transmission spectra as shown in
Figs.~2($\mathrm{b_2}$-$\mathrm{b_3}$). In this case, the photon can not be transmitted from ports 2
and 3 to ports 4 and 1 , respectively. Due to the large frequency detuning,
the photon entering from port 2 can not be transmitted to the CCW mode of
the $f$ resonator. It also cannot couple with the CW mode of the $f$
resonator, as the direction of photon transmission is opposite to it. So
that the photon from port 2 transmits through the $a$ waveguide directly to
port 1. Also, because of the large frequency detuning, the photon entering
from port 3 can not be transmitted to the CW mode of the $d$ resonator. It
also cannot couple with the CCW mode of the $d$ resonator, as the direction
of photon transmission is opposite to it. So that the photon from port 3
transmits through the $b$ waveguide directly to port 4. In contrast, the
photon entering from port 1 (4) couples to the CW (CCW) travelling mode with
the couping $g_{a} (g_{b})$, so that the photon can be transferred to port 3
(2). Therefore, four frequency point photon circulators can be achieved in
this system.

The circulator can also be in the direction of a positive $8$ character,
i.e. $1\rightarrow 2\rightarrow 4\rightarrow 3\rightarrow 1$, at $\delta
=\pm \sqrt{g_{a}^{2}g_{b}^{2}+4v_{g}^{2}J^{2}+v_{g}^{2} \Delta_{F1} ^{2}}%
/(2v_{g})+\Delta _{F1}/2$, see Fig.\,\ref{fig2}($\mathrm{c_1}$) (The position marked by a green
star). The routing behavior can be understood in a similar way as given
above. The photon can not be transmitted from ports 1 and 4 to ports 4 and 2
, respectively. Due to the large frequency detuning, the photon entering
from port 1 can not be transmitted to the CW mode of the $f$ resonator. It
also cannot couple with the CCW mode of the $f$ resonator, as the direction
of photon transmission is opposite to it. So that the photon from port 1
transmits through the $a$ waveguide directly to port 2. The photon entering
from port 4 can not be transmitted to the CCW mode of the $d$ resonator. It
also cannot couple with the CW mode of the $d$ resonator, as the direction
of photon transmission is opposite to it. So that the photon from port 4
transmits through the $b$ waveguide directly to port 3. In contrast, the
photon entering from port 2 (3) couples to the CCW (CW) travelling mode with
the couping $g_{a}$ $(g_{b})$, so that the photon can be transferred to port 4
(1). In addition, when $\Omega_{1}<0$ and $\Omega_{2}=0$ (or $\Omega_{1}=0$
and $\Omega_{2}>0$), we have plotted the transmission probability
distribution as a function of detuning $\delta$, as shown in Fig.\,\ref{fig2}($\mathrm{c_1}$-$\mathrm{c_3}$). We
found that Fig.\,\ref{fig2}($\mathrm{c_2}$-$\mathrm{c_3}$) is mirror symmetric in Fig.\,\ref{fig2}($\mathrm{b_2}$-$\mathrm{b_3}$) with $\delta=0$ as
the mirror surface.

\begin{figure}[tbp]
	\centering\includegraphics[width=0.9\textwidth]{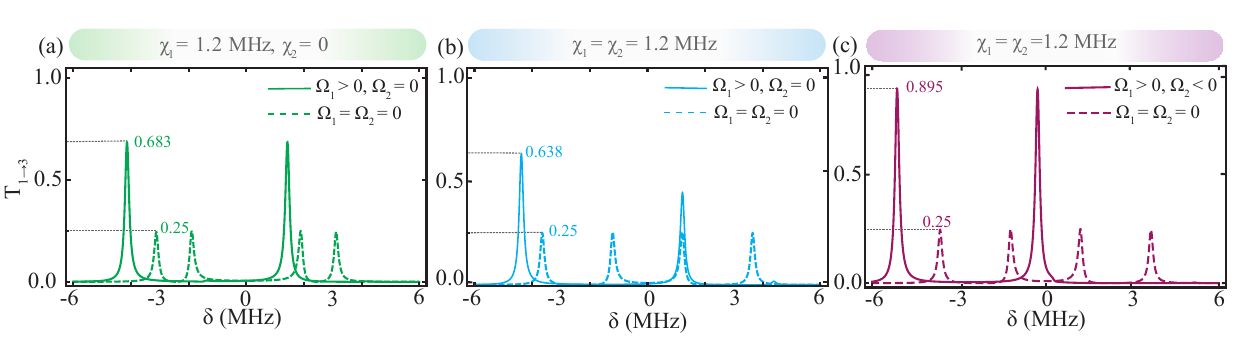}
	\caption{(Color online) Transmission probability distribution as a function
		of detuning $\protect\delta $ for two resonators . (a) $\Omega _{1}=29$ kHz,
		$\Omega _{2}=0$ (blue solid line), $\Omega _{1}=\Omega _{2}=0$ (blue dashed
		line). (b) $\Omega _{1}=29$ kHz, $\Omega _{2}=-29$ kHz(purple dashed line). Other parameters can be
		found in the main text. }
	\label{fig4}
\end{figure}

Then, we consider the photon transmission spectra of two resonators rotating
in the same directions (e.g., $\Omega_{1}=\Omega_{2}>0$), as shown in the
Figs.\,\ref{fig3}(a-b). Consistent with our previous calculation results, the
probability of phohton transmission from any port is $1/2$, and the photon
transmission is reciprocal in the same waveguide. In Figs.\,\ref{fig3}(c-f), we have
plotted the photon transmission spectra of two resonator rotating in
opposite directions, e.g., $\Omega_{1}>0$, $\Omega_{2}<0$, and $%
|\Omega_{1}|=|\Omega_{2}|$ or $\Omega_{1}<0$, $\Omega_{2}>0$, and $%
|\Omega_{1}|=|\Omega_{2}|$. Here, we found that a frequency point $\delta
=\pm\sqrt{g_{a}^{2}g_{b}^{2}+4v_{g}^{2}J^{2}+v_{g}^{2}\left( \Delta
_{F1}+\Delta_{F2}\right) ^{2}}/(2v_{g})+(\Delta _{F2}-\Delta _{F1})/2$ ($%
\delta = \pm \sqrt{g_{a}^{2}g_{b}^{2}+4v_{g}^{2}J^{2}+v_{g}^{2}\left( \Delta
_{F1}+\Delta_{F2}\right) ^{2}}/(2v_{g})+(\Delta _{F2}+\Delta _{F1})/2$) can
achieve an reverse or positive $8$ circulator, which means to realize the
two photon circulator frequency point. Besides, we also found that at a
specific angular velocity $|\Omega_{1}|=|\Omega_{2}|= M/G$ ($M=\sqrt{%
g_{a}^{2}g_{b}^{2}+4v_g^{2}J^{2}}/(2v_{g})$, $G=R \omega _{c}\left(
n^{2}-1\right)/(cn)$), there was one frequency point that the photon
entering from any waveguide can be completely routed to the other waveguide.
We also found that Figs.\,\ref{fig3}(c-d) is mirror symmetric in Figs.\,\ref{fig3}(e-f) with $%
\delta=0$ as the mirror surface.

Comparad to a single resonator case, two resonators can have four frequency
points for photon circulators, due to the coupling between resonators caused
energy level splitting and the Fizeau drag induced splitting of the
resonance frequencies of the two counter-travelling optical modes. In
addition, we also found that there was one frequency point that the photon
entering from any waveguide can be completely routed to the other waveguide.
The direction of the photon circulator has also changed, form clockwise
(counterclockwise) to reverse (positive) 8-word. The different directions of
the photon circulator are due to the coupling between the CW (CCW) mode of
the $f$ resonator and the CCW (CW) mode of the $d$ resonator.


\section{\label{Sec:4}Backscattering Effect}

The WGM resonators has two degenerate modes corresponding to CW and CCW
modes. In practical devices, backscattering is unavoidable. They may be
coupled together if the surface of the resonator is rough or the material is
uneven\cite{zhu46,Guo2023,Jiao2020}. So the Hamiltonian of the system is
added with a term
\[
\begin{aligned} \hat{H}_{bs}=\chi _{1}\hat{f}_{cw}^{\dagger
}\hat{f}_{ccw}+\chi _{2}\hat{d}_{ccw}^{\dagger }\hat{d} _{cw}+\mathrm{H.c.}
\end{aligned}
\]%
where $\chi _{1}$ ($\chi _{2}$) is the coupling strength of CW and CCW modes
inside the $f$ $(d)$ resonator due to backscatter. Using the same method as in the previous
section, we can calculate the probability $T_{i\rightarrow j}$ for each
port. Figure\thinspace \ref{fig4} shows the probability of the photon
transitioning from port 1 to port 3 under different conditions.

\begin{figure}[tbp]
\centering\includegraphics[width=0.7\textwidth]{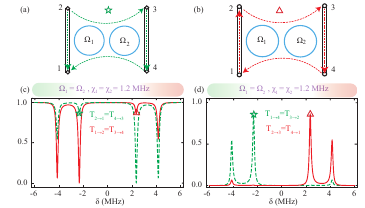}
\caption{(Color online) The routing behaviors of the circulator for two
resonators (a-b). Transmission probability distribution as a function of
detuning $\protect\delta $ for two resonators (c-d). (c-d) $\Omega _{1}=\Omega
_{2}=24$ kHz. Other parameters can be found in the main text.}
\label{fig5}
\end{figure}
In Fig.\,\ref{fig4}(a), the dashed line indicates no rotation, but the $f$ resonator
has backscattering, e.g. $\Omega _{1}=\Omega _{2}=0,\chi _{1}=1.2$ MHz, $\chi _{2}=0$%
. The solid line indicates rotation of the $f$ resonator under the condition
of the dashes line, e.g. $\Omega _{1}>0,\Omega _{2}=0,\chi _{1}=1.2$ MHz, $\chi _{2}=0
$. We found that the maximum value of $T_{1\rightarrow 3}$ increased from
0.25 to 0.683. So the results indicate that rotation is robust against
backscattering. We also found that the frequency points that can be
transmitted have shifted to the left, which is caused by the clockwise
rotation of the resonator. On the contrary, when the resonator rotates
counterclockwise, the transmission frequency point shift to the right. In
Fig.~4(b), the dashed line indicates that both resonators are static but
have backscattering, e.g. $\Omega _{1}=\Omega _{2}=0$, $\chi _{1}=\chi _{2}=1.2
$ MHz. In contrast to Figure 4(a) a resonator has backscatter, we found that adding backscattering in another
resonator does not change the value of $T_{1\rightarrow 3}=0.25$, bacause
the four ports are transmitted with equal paobability. In the case of
backscatter in both resonators, the soild line indicates tthe $f$ resonator
rotates clockwise and the $d$ resonator static, e.g. $\Omega _{1}>0$, $%
\Omega _{2}=0$, $\chi _{1}=\chi _{2}=1.2$ MHz. The
maximum value of $T_{1\rightarrow 3}$ has decreased from 0.683 to 0.638.
This is cauesd by the backscatter of the $d$ resonator. When the $f$
resonator rotates clockwise and the $d$ resonator rotates counterclockwise ($%
\Omega _{1}>0$, $\Omega _{2}<0$, $|\Omega _{1}|=|\Omega _{2}|$), the
transmission probability $T_{1\rightarrow 3}$ can reach 0.895, as shown in
Fig.\,\ref{fig4}(c). In addition, if $\Omega _{1}<0$, $\Omega _{2}>0$, $|\Omega
_{1}|=|\Omega _{2}|$, its transmission spectrum can be obtained in Fig.\,\ref{fig4}(b)
though mirror reflection symmetry with $\delta =0$ as the mirror surface.

In the previous section, when the rotational angular velocities of the two
resonators are equal and the direction is the same, the photon transmission
is reciprocal in the same waveguide. Here,  due to the addition of
backscattering, we have broken this reciprocity. In Fig.\,\ref{fig5}, the transmission
spectra of photons are plotted by adding backscattering to the resonators
when the rotational angular velocities of the two resonators are equal and
the direction is the same. We found two frequency points that can implement
a single photon circulator, and the routing behavior is shown in
Figs.\,\ref{fig5}(a-b). The circulator can be seen from the transmission spectra as
shown in Figs.\,\ref{fig5}(c-d). Unlike the 8-word circulator, the direction of the
circulator is clockwise or counterclockwise circulator.


\section{\label{Sec:5}Conclusion}

We studied the coherent transmission of a single photon in a coupled system
of two rotating WGM resonators and two waveguides. This system could form a
four-port circulator, whose routing direction depended on the frequency of
the incident photon, and the direction of rotation of the two resonators. We found that when both resonators
rotated at identical angular velocities, single-photon transmission
exhibited reciprocity within the same waveguide. When the angular velocities
differed, usually there were four frequency points capable of
facilitating single-photon circulation. However, when two resonators
rotated in opposite directions with equal angular velocities, besides photon circulates at two frequency points, there is one frequency point at which a single photon input from any waveguide is completely routed to another. When the backscattering of a resonator is present, the single-photon circulation suppressed by the internal defect-induced backscattering can be restored by rotating the resonator, which is witnessed by enhancement of the probability of a single photon transfer.

We note that the presence of  the reciprocity  shut down the circulator for photons with arbitrary frequency. To turn on the circulator, the resonator must rotate at different angular velocities, so that the reciprocity of photon transmission is broken.  Furthermore, one can tune the angular velocity to turn on or off circulating  photons with a given frequency. The nonreciprocity of quantum circulators can be used to design novel quantum photonic device, which may have extensive applications in future quantum technologies.

\begin{backmatter}
		\bmsection{Funding}
	This work was supported by NSFC Grants No.11935006, No. 12075082, No. 12247105, No. 12205054, the science and technology innovation Program of Hunan Province (Grant No. 2020RC4047), National Key R $\&$ D Program of China (No. 2024YFE0102400), the Hunan Provincial major Sci-Tech Program (2023ZJ1010) and PH. D. Research Foundation (BSJJ202122).
		\bmsection{Disclosures}
		The authors declare no conflicts of interest.
		
\end{backmatter}


\end{document}